# The Representation of Quantum Field Theory without Infinity Bare Masses and Coupling Constants of Fermions

V.P.Neznamov[1]

RFNC-VNIIEF, Mira pr., 37, Sarov, Russia, 607188

## Abstract

The paper presents the representation of quantum field theory without introduction of infinity bare masses and coupling constants of fermions. Counter-terms, compensating for divergent quantities in self-energy diagrams of fermions and vacuum polarization diagrams at all orders of the perturbation theory, appear in the appropriate Hamiltonians under the special time-dependent unitary transformation.

---

[1] E-mail: neznamov@vniief.ru

## 1. Introduction

It is well known that in Quantum Field Theory, when calculating matrix elements, containing Feynman diagrams with closed loops of fermion and boson lines, divergent quantities emerge. A special procedure has been developed to select finite expressions in renormalization theories. Renormalization of masses and coupling constants of elementary particles are available therein. In the renormalization procedure, it is assumed that masses and coupling constants of elementary particles observed experimentally consist of bare and additional parts

$$m_{phys} = m_0 + \Delta m \tag{1}$$
$$q_{phys} = q_0 + \Delta q \tag{2}$$

Each part ($m_0$ and $\Delta m$; $q_0$ and $\Delta q$) is infinite but in sum they are finite and equal to the values $m_{phys}$ and $q_{phys}$ observed in the experiments. This mysterious procedure in conjunction with two other renormalization constants $Z_1 = Z_2$ allows elimination of infinite expressions at all orders of the perturbation theory and enables unprecedented accurate calculations of physical values obtained in the experiments. The in-depth description of the procedure of renormalization is provided in monographs and textbooks on quantum field theory (see, for instance, [1] - [5]).

In the present paper, the author proposes to renounce the idea of dividing masses and coupling constants of fermions into two infinite parts and to deal with finite $m_{phys}$ and $q_{phys}$. The infinities will be compensated for at all the orders of the perturbation theory due to the energy shifts in the Hamiltonians caused by special time-dependent unitary transformation.

Quantum electrodynamics (QED) formulated in the Hamiltonian form will be used below as an example of examination. In this case, $m_{phys} = m$ is the electron (or positron) mass; $q_{phys} = e$ is the electron charge.

Section 2 of the paper describes some of the properties of the time-dependent unitary transformations. Section 3 explores the replacing procedure of renormalization with infinity bare mass and charge of electron in self-energy and vacuum polarization diagrams. In the Conclusions, the obtained results will be discussed.

The system of units $\hbar = c = 1$ will be used below; The Minkowsky space metric is taken in the form of [^2] $g^{\mu\nu} = diag[1,-1,-1,-1]$; $x = (t, \mathbf{x})$; $p^{\mu} = i\dfrac{\partial}{\partial x_{\mu}}$.

## 2. Time-dependent unitary transformations

Firstly, let us examine the quantum mechanics of an electron interacting with the electromagnetic field.

In this case, the Dirac equation can be written in the following form

$$p_0 \psi_D = H_D \psi_D = (\boldsymbol{\alpha}\mathbf{p} + \beta m + e\alpha_{\mu}A^{\mu})\psi_D, \qquad (3)$$

where $\psi_D(x)$ is the four-component wave function of the electron; $A^{\mu}(x)$ - is the four-vector of the electromagnetic field; $H_D$ is the Dirac Hamiltonian; $\alpha^i, \beta$ are the Dirac matrices;

$\alpha^{\mu} = \begin{cases} 1, & \mu = 0 \\ \alpha^i, & \mu = i = 1,2,3 \end{cases}$ ; $e$ -is the electron charge.

Let $R(t)$ be some time-dependent unitary transformation of the wave function $\psi_D(x)$. Then,

$$\psi_R(x) = R(t)\psi_D(x). \qquad (4)$$

The Dirac equation (3) is transformed to the form:

$$p_0 \psi_R(x) = H_R \psi_R(x), \qquad (5)$$

where

$$H_R = R(t) H_D R^{+}(t) - iR(t)\dfrac{\partial R^{+}}{\partial t}. \qquad (6)$$

All the other transformed operators $O_R$ are equal to

$$O_R = R O R^{+}. \qquad (7)$$

The equation (5) is equivalent to the initial equation (3).
Indeed,

$$p_0 \psi_R(x) = p_0 (R\psi_D(x)) = H_R R \psi_D(x),$$

$$Rp_0\psi_D(x) + \left(i\dfrac{\partial}{\partial t}R\right)\psi_D(x) = RH_D R^{+} R\psi_D(x) - iR\dfrac{\partial R^{+}}{\partial t}R\psi_D(x).$$

[^2]: The Greek letters take on values of 0, 1, 2, 3 and the Roman letters take on values of 1, 2, 3.

By multiplying, on the left-hand side, by the operator $R^+$ and taking into account that $\frac{\partial R^+}{\partial t}R = -R^+\frac{\partial R}{\partial t}$, we will derive the initial Dirac equation (3).

Then, let us consider the Hamiltonian of unquantized electron-positron fields interacting with classical electromagnetic fields:

$$\mathcal{H} = \int \psi_D^+(x) H_D \psi(x) d\mathbf{x} = \int \psi_D^+(x)\left(\boldsymbol{\alpha}\mathbf{p} + \beta m + e\alpha_\mu A^\mu(x)\right)\psi_D(x) d\mathbf{x}. \tag{8}$$

In the expression (8), the Hamiltonian of free electromagnetic fields is omitted; the sign "+" means the Hermitian conjugation.

Let us apply the unitary transformation $R(t)$ to the $\psi_D(x), \psi_D^+(x)$ fields

$$\psi_R(x) = R(t)\psi_D(x); \quad \psi_R^+(x) = \psi_D^+(x) R^+(t). \tag{9}$$

The Hamiltonian (8) in the $R$-representation is

$$\mathcal{H}_R = \int \psi_R^+(x) H_R \psi_R(x) d\mathbf{x}. \tag{10}$$

In the Hamiltonian (10), let us proceed to the record via the fields $\psi_D(x), \psi_D^+(x)$. Then, taking into account (9) and (6), we will have

$$\begin{aligned}\mathcal{H}_R &= \int \psi_D^+(x) R^+\left(RH_D R^+ - iR\frac{\partial R^+}{\partial t}\right) R\psi_D(x) d\mathbf{x} = \\ &= \int\left[\psi_D^+(x) H_D \psi_D(x) - \psi_D^+(x) i\frac{\partial R^+}{\partial t} R\psi_D(x)\right] d\mathbf{x}.\end{aligned} \tag{11}$$

The Hamiltonian (11) differs in the energy shift $-i\frac{\partial R^+}{\partial t}R$ from the initial Hamiltonian (8) This shift can be used to replace the standard procedures of renormalization of mass and charge of electron (positron) in quantum electrodynamics.

Let us note that the Lagrangian of unquantized electron-positron and electromagnetic fields

$$\mathcal{L} = \int \psi_D^+ i \frac{\partial \psi_D}{\partial t} d\mathbf{x} - \mathcal{H} \tag{12}$$

is invariant to the transformation (4).

# 3. Divergent expression compensation in self-energy diagrams of an electron and a photon

Let us select the unitary transformation (4) in the following form:

$$R(t) = T - \exp\left(i\beta\omega_m t + i\omega_e e \int \alpha_\mu A^\mu(x) dt\right), \quad (13)$$

where $T - \exp$ is the Feynman-introduced function with ordering of operator factors in the Taylor expansion [6], [7]; $\omega_m, \omega_e$ are arbitrary numeric parameters, which can be arbitrary large. We will assume that $\omega_m, \omega_e$ depend on the coupling constant $e$.

The Hamiltonian of classical fields (11) in the $R$-representation for the transformation (13) becomes equal to

$$\mathcal{H}_R = \int \psi_D^+(x)\left(\boldsymbol{\alpha}\mathbf{p} + \beta m + e\alpha_\mu A^\mu(x) - \beta\omega_m - \omega_e e\alpha_\mu A^\mu(x)\right)\psi_D(x) d\mathbf{x}. \quad (14)$$

The Hamiltonian (14) can be written as the sum of the free field Hamiltonian $H_0$ and the interaction Hamiltonian $H_{int}$.

$$H_0 = \int \psi_D^+(x)(\boldsymbol{\alpha}\mathbf{p} + \beta m)\psi_D(x) d\mathbf{x}, \quad (15)$$

$$H_{int} = \int \psi_D^+(x)\left(e\alpha_\mu A^\mu(x) - \beta\omega_m - \omega_e e\alpha_\mu A^\mu(x)\right)\psi_D(x) d\mathbf{x}. \quad (16)$$

Then, taking into account the division of (14) into (15) and (16), one can perform quantization of electron-positron and electromagnetic fields in a standard way and proceed to quantum electrodynamics.

If, in the terminology [5], to determine $-i\Sigma(p)$ as a sum of all one-particle irreducible (1PI) diagrams with two external fermion lines, then the counter-term with $\Delta m$ in the QED interaction Hamiltonian resulting from splitting of (1) and compensating for the divergent expressions in the self-energy parts of the Feynman diagrams at all orders of the perturbation theory is equal to:

$$\Delta m = \Sigma(p^2 = m^2). \quad (17)$$

At the lowest (second) order of the perturbation theory

$$\Delta m^{(2)} = \Sigma^{(2)}(p^2 = m^2). \quad (18)$$

It is obvious that if in the Hamiltonian (16) the parameter $\omega_m$ is taken to be

$$\omega_m = \Delta m, \quad (19)$$

then the similar compensation for the divergent expressions in the self-energy parts of the Feynman diagrams will occur due to the unitary transformation (13) without splitting the electron mass into the bare and additional parts.

In the existing quantum field theory, there are the notions of bare and physical (observed) coupling constants of elementary particles (see the equality (2)). In quantum electrodynamics, the relationship between them is determined by the expression

$$e = e_0 Z_3^{1/2}. \tag{20}$$

At the finite charge of electron observed in the experiments, the bare charge $e_0$ and the value $Z_3$ are infinitely large.

If we define

$$i\Pi^{\mu\nu}(q) = \left(q^2 g^{\mu\nu} - q^\mu q^\nu\right)\Pi(q^2) \tag{21}$$

as the sum of all OPI insertions into the photon propagator, then the renormalization constant $Z_3$ is

$$Z_3 = \frac{1}{1-\Pi(0)} = 1 + \Pi^{(2)}(0) + \Pi^{(4)}(0) + ... \tag{22}$$

In (22), $q^\mu$ is a photon four-impulse

In our case, the last summand in the Hamiltonian (16) can play the role of the constant $Z_3$ if we assume that

$$\omega_e = 1 - \left(1-\Pi(0)\right)^{-1/2}. \tag{23}$$

Then, this summand as well as $Z_3$, will compensate for infinitely large quantities, emerging when calculating 1PI diagrams of vacuum polarization, at all the orders of the perturbation theory. In this case, there is no need for dividing the physically observed charge of the electron into infinitely large bare and additional parts.

## 5. Conclusions

In the paper, QED is used as an example to show that when using a time-dependent unitary transformation generating energy shifts in the appropriate Hamiltonians of interacting quantum fields, it is possible to compensate the divergent expressions without introduction of infinity bare masses and coupling constants of fermions at all the orders of the perturbation theory. The only renormalization constant $Z_1 = Z_2$ remains in the theory. The obtained results are also applicable to other interactions of the Standard Model formulated in the Hamiltonian form.

The energy shift $\omega_m$ in the Hamiltonian (14) is not physically observable (see the relevant discussion in [8]). On the contrary, the shift $\omega_e$ in (14) changes the constant of the electromagnetic interaction and should be selected based on the condition of agreement with the experiment. In this paper, it is initially assumed to use masses and coupling constants of half-spin particles observed experimentally.

Surely, the problem of calculations of finite values for mass operator of fermions $\Sigma(p^2 = m^2)$ and polarization operator $\Pi(q = 0)$ persists in the theory as before. It is not inconceivable that the problem could be solved through selection of some unitary transformation $R_1(t)$. Application of this transformation will lead to compensation for logarithmically divergent values in the self-energy diagrams of fermions and vacuum polarization diagrams and the remaining finite values will determine the sought values. The author believes that the retrieval of such a transformation is topical.